\documentclass[9pt,journal,compsoc]{IEEEtran}
\usepackage{tikz}
\usepackage{tikz-cd}
\usetikzlibrary{arrows}
\usepackage{tikz-qtree}

\ifCLASSOPTIONcompsoc
  \usepackage[nocompress]{cite}
\else
  \usepackage{cite}
\fi
\usepackage{listings}

\usepackage{changes}
\definechangesauthor[color=orange]{rs}
\definechangesauthor[color=red]{is}

\ifCLASSINFOpdf
\else
\fi
\usepackage{algorithm}
\usepackage[noend]{algpseudocode}
\usepackage{url}

\usetikzlibrary{shapes,positioning,chains,fit,calc}
\usetikzlibrary{decorations.markings}
\tikzstyle arrowstyle=[scale=1]
\tikzstyle directed=[postaction={decorate,decoration={markings,
		mark=at position .65 with {\arrow[arrowstyle]{stealth}}}}]
\tikzstyle reverse directed=[postaction={decorate,decoration={markings,
		mark=at position .65 with {\arrowreversed[arrowstyle]{stealth};}}}]
\begin{document}
%
\title{Community Detection Across Emerging Quantum Architectures}

%
%
%
%

\author{Ruslan~Shaydulin,
        Hayato~Ushijima-Mwesigwa,
        Ilya~Safro,
        Susan~Mniszewski,
        Yuri~Alexeev%
\IEEEcompsocitemizethanks{\IEEEcompsocthanksitem Ruslan~Shaydulin, Hayato~Ushijima-Mwesigwa and Ilya~Safro are with School of Computing, Clemson University, Clemson, SC 29634. E-mail: rshaydu@g.clemson.edu
\IEEEcompsocthanksitem Susan~Mniszewski is with Computer, Computational, \& Statistical Sciences Division, Los Alamos National Laboratory, Los Alamos, NM 87545.
\IEEEcompsocthanksitem Yuri~Alexeev is with Computational Science and Leadership Computing Divisions, Argonne National Laboratory, Argonne, IL 60439}
}

\IEEEtitleabstractindextext{%
\begin{abstract}
One of the roadmap plans for quantum computers is an integration within HPC ecosystems assigning them a role of accelerators for a variety of computationally hard tasks. However, in the near term, quantum hardware will be in a constant state of change. Heading towards solving real-world problems, we advocate development of portable, architecture-agnostic hybrid quantum-classical frameworks and demonstrate one for the community detection problem evaluated using quantum annealing and gate-based universal quantum computation paradigms.

\end{abstract}
}


\maketitle

\IEEEdisplaynontitleabstractindextext

%
\IEEEpeerreviewmaketitle

\section{Introduction}

During the last two years a race of industrial and research organizations has been opened to develop a ready-to-implement engineering solution for quantum computing (QC). It resulted in the QC market closely resembling the ascent ages of classical computing industry.
Namely, there were many underdeveloped computing architectures which being incompatible with each other required significant efforts in porting software and algorithmic solutions between them. Given a broadly supported opinion that in the near term we are unlikely to become witnesses to flexible large-scale quantum architectures, there is a critical need to develop  portable, architecture-agnostic hybrid quantum-classical frameworks that will allow solving large-scale computational problems on small-scale quantum architectures.

There are multiple emerging quantum computation paradigms. The performance comparison of these paradigms is an important research topic. In this paper, we present for the first time a performance comparison of two leading quantum computation paradigms - D-Wave  quantum annealing and gate-based universal quantum computation. Both approaches have great potential for achieving quantum speedup for a number of important  problems~\cite{king2018observation,romero2018strategies,ambainis2018quantum,dunjko2018computational}. 



The first approach, quantum annealing (QA), is based on adiabatic quantum computation (AQC)~\cite{mcgeoch2014adiabatic}. QA solves computational problems by using a guided quantum evolution~\cite{yang2017optimizing}. 
The evolution starts with an initial Hamiltonian with an easy-to-prepare ground state and ends up in the ground state of the problem Hamiltonian. QA is based on the adiabatic theorem that guarantees that if the Hamiltonian is evolved slowly then transitions to excited states are suppressed during the adiabatic evolution~\cite{yang2017optimizing}.
The D-Wave quantum annealer uses superconducting flux qubits \cite{amin2004,dwave2018} and has been shown to solve optimization problems on graphs \cite{ushijima2017graph}, machine learning \cite{omalley2017}, traffic flow optimization \cite{neukart2017}, and simulation problems \cite{harris2018}.
Quantum and hybrid quantum-classical approaches have been employed.

The second approach is often referred to as the gate-based or universal QC. This mode of QC was theoretically demonstrated to have a great potential for exponential speedups over best known classical algorithms~\cite{nielsen2002quantum}. In the near term, the capability of the quantum devices is limited by the number of qubits, low fidelity of gates, and lack of error correction. These limitations constrain us to using low-depth quantum circuits (i.e., quantum circuits with few gates) on a small number of qubits. Within the constraints of near-term intermediate-scale quantum (NISQ) technology~\cite{preskill2018quantum}, a number of hybrid quantum-classical algorithms were developed and experimentally demonstrated to solve small problems. One of the most promising of such algorithms is Quantum Approximate Optimization Algorithm (QAOA)~\cite{farhi2014quantum,farhi2016quantum}. QAOA is inspired by adiabatic quantum computation. Similarly to AQC and QA, the evolution path starts with an easy-to-prepare Hamiltonian in the ground state and evolves to the final Hamiltonian that encodes the solution of the problem remaining in the ground state. However, unlike QA in QAOA the evolution is performed by applying a series of parametrized gates called ansatz~\cite{mcclean2016theory} which is parametrized by a set of variational parameters. This is accomplished by a hybrid approach that combines quantum evolution and classical variational optimization for optimal QAOA parameters~\cite{yang2017optimizing} with the goal of finding the evolution path that prepares the ground state of the problem Hamiltonian. 

\section{Methodology}

This work addresses three main challenges. First, we show how to use quantum computing to solve the community detection problem, a well known NP-hard problem. Second, we present an approach to solving realistic large problems using the NISQ hardware with a limited number of noisy qubits.  Third, we demonstrate a method that is portable across two leading quantum computation paradigms and can be easily extended to future hardware.

The community detection problem (or modularity graph clustering) has a variety of applications ranging from biology to social network analysis~\cite{palla2005uncovering,su2010glay,bardella2016hierarchical,nicolini2017community}. 
Its complexity \cite{brandes2006maximizing} and practical importance justify an attempt to solve it using QC. The goal of the community detection is to split nodes of a graph $G=(V,E)$ into communities by maximizing its modularity~\cite{2006PNAS..103.8577N}:

\begin{equation}
H = \frac{1}{4|E|}\Sigma_{ij}(A_{ij} - \frac{k_ik_j}{2|E|})s_is_j =  \frac{1}{4|E|}\Sigma_{ij}B_{ij}s_is_j,
\label{eq:mod}
\end{equation}

\noindent where $s_i\in\{-1,+1\}$ are variables indicating node $i$th community assignment, $k_i$ is a degree $i\in V$, and $A$ is the adjacency matrix of $G$. 
In this paper, we will focus on clustering the graph into two communities. There are several approaches to generalize the problem for cases when the number of communities is greater than 2.

The clustering of large networks is currently impossible with existing quantum computers because of the small number of available qubits.
This limitation applies both to quantum annealing~\cite{ushijima2017graph} and universal quantum computing~\cite{otterbach2017unsupervised}. To tackle large problems using available quantum hardware, we use a hybrid quantum-classical local-search approach. Our approach is inspired by existing numerous local-search heuristics (see~\cite{rotta2011multilevel} for a review). Our algorithm finds a solution to the global community detection problem by selecting subproblems small enough to fit on the target quantum computer, solving them using a quantum algorithm and iterating until the solution to the global problem is found. The outline is presented in Algorithm~\ref{alg:outline}.

\begin{algorithm}
 \caption{Community Detection}\label{alg:outline}
\begin{algorithmic}
\Procedure{Community detection}{Graph $G$}
\State solution = initial\_guess($G$)
 \While{not converged}
  	\State $X$ = populate\_subset($G$)
  	\State // \textit{using QAOA or D-Wave QA}
  	\State candidate = solve\_subproblem($G$, $X$)
  	\If{$\mbox{candidate} > \mbox{solution}$}
  	\State solution = candidate
    \EndIf
 \EndWhile
 \EndProcedure
\end{algorithmic}
\end{algorithm}

In particular, we start with a random community assignment. At each step we select a subproblem (subset of vertices $X\subset V$) by taking the vertices with highest potential gain if moving them from one community to another. The gain for each vertex can be computed efficiently~\cite{2006PNAS..103.8577N}. Then we fix the community assignment of all $i\not\in X$, encode them into the problem as boundary conditions (denoted by $\tilde{s}_j$, a typical technique in many heuristics \cite{leyffer2013fast,hager2018multilevel}) and maximize

\begin{equation}
\label{eq:subproblem}
\arraycolsep=1.4pt
\begin{array}{r c l}
Q_{s} & = & \sum_{i>j | i,j\in X}2B_{ij}s_is_j +  \sum_{i\in X}\sum_{j\not\in X}2B_{ij}s_i\tilde{s}_j \\
 & = & \sum_{i>j| i,j\in X}2B_{ij}s_is_j + \sum_{i\in X}C_{i}s_i. 
\end{array}
\end{equation}

The subproblems are solved using QC. To satisfy the constraints of available hardware, we fix the subproblem size to some small number (in our experiments, it was 25). 


\section{Implementation details and Results}

We implement our local search algorithm in Python using the graph methods provided by NetworkX~\cite{hagberg2008}. The novelty of our approach is that it allows to use D-Wave QA, QAOA and classical Gurobi \cite{optimization2014inc} solvers interchangeably simply by passing different flags, enabling rapid prototyping and direct comparison of different methods as the hardware and its capabilities evolve. Additionally, Gurobi was used as a global optimization solver for the sake of quality comparison. To our knowledge this is the first attempt to directly compare universal quantum computing and quantum annealing. Our framework is also easily extendable, making it possible for researchers to add new backends as they become available. We plan to release the framework as an open-source project.

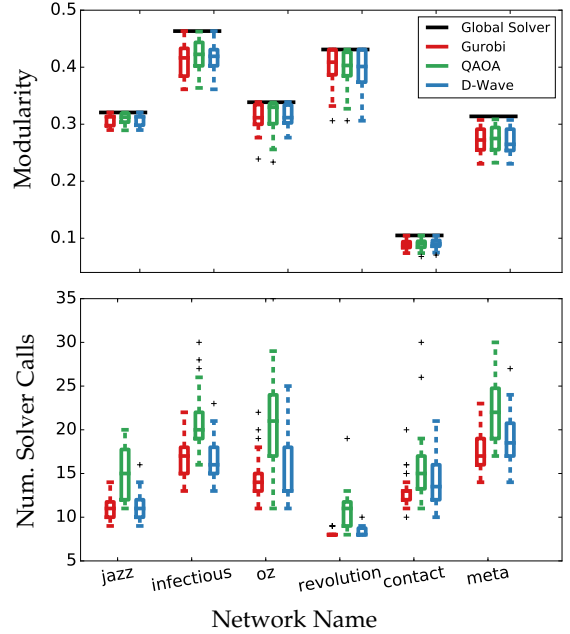
\begin{figure}[htb]
 \begin{tikzpicture} 
  \node (img)  {\includegraphics[width=0.8\linewidth]{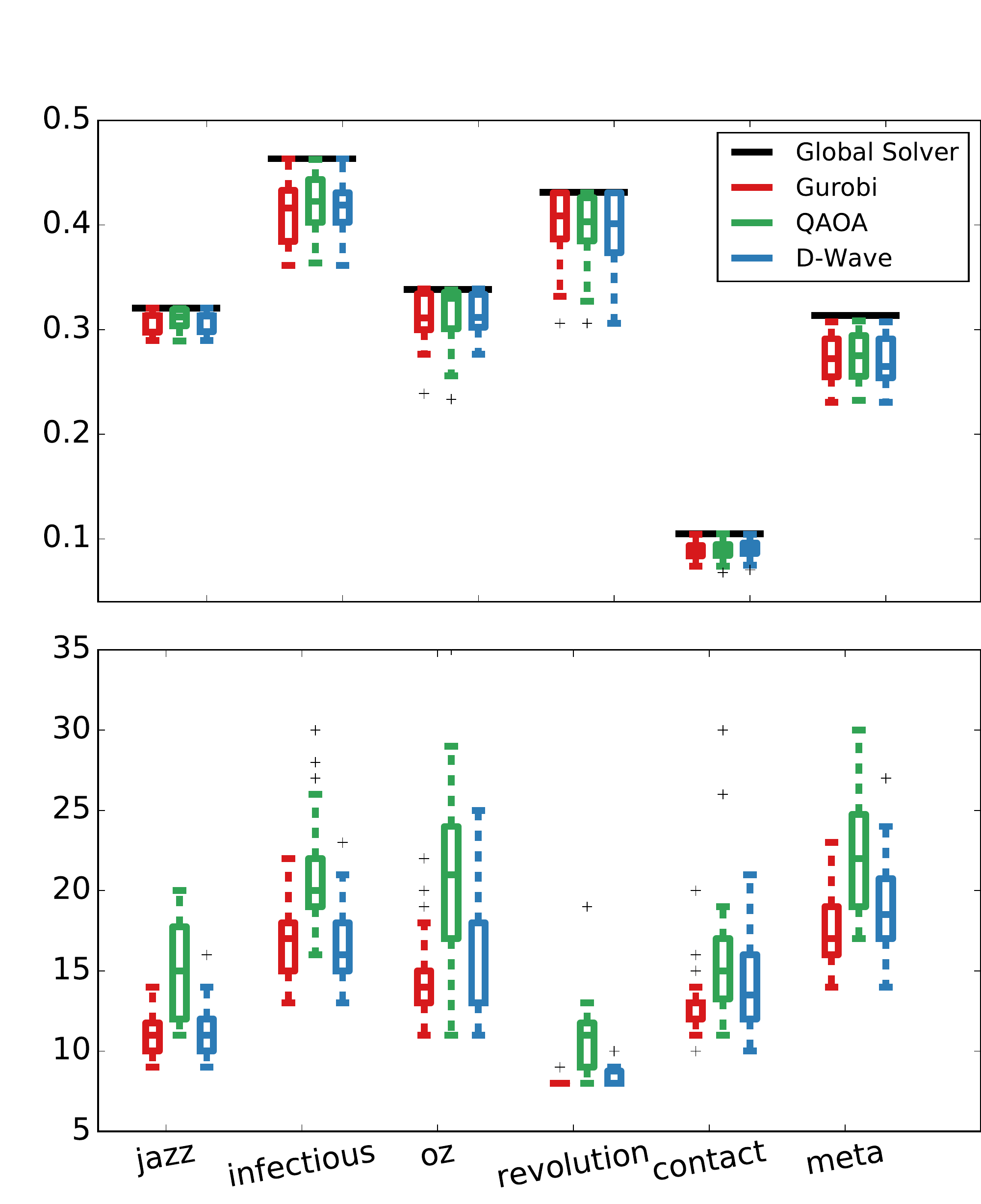}};
  \node (temp) [left=of img]{};
  \node[below=of img, node distance=0cm, yshift=1.1cm,font=\color{black}] {Network Name};
 \node[below=of temp, node distance=2cm, anchor=center,yshift=-1.0cm,,xshift=1.2cm, rotate=90,font=\color{black}] {Num. Solver Calls};
  \node[above=of temp, node distance=2cm, xshift=1.2cm,yshift=0.8cm, rotate=90, anchor=center,font=\color{black}] {Modularity};
 \end{tikzpicture}
  \vspace{-0.4cm}
 \caption{Box-plots comparing modularity scores (greater is better) and number of solver calls (less is better) respectively for the three different solvers. For the graph {\tt oz}, Gurobi and D-Wave returned a modularity score greater than the Global Solver (best known value)}
 \label{fig:results}
\end{figure}

Our results are presented in Figure \ref{fig:results}. In these experiments, we used the Intel-QS~\cite{smelyanskiy2016qhipster} simulator for QAOA (at the time our group did not have access to a universal quantum computer of sufficient size). We use six real-world networks from the KONECT dataset~\cite{kunegis2013konect} with up to 400 nodes as our benchmark. For each network, we ran 30 experiments with different random seeds. The same set of seeds is used between three backend solvers, making the results directly comparable. The subproblem size is fixed at 25 (i.e., 25 qubits are used). Our results demonstrate that the quantum local search approach with both quantum methods is capable of achieving results comparable to state-of-the-art, with a potential to outperform as hardware evolves.

 



 \vspace{-0.78cm}
 
 \vspace{-0.6cm}
\section{Discussion}




In the near term, quantum hardware will be in a constant state of change. Many different NISQ-era hardware solutions will appear and some will be abandoned. In the midst of such evolutionary times, we want to be able to continue research in quantum algorithms and head towards solving real-world problems. To accomplish this, we need portable, architecture-agnostic hybrid quantum-classical frameworks that will allow solving large-scale computational problems on small-scale quantum architectures. Moreover, these frameworks need to be robust and future-proof. In this work, we have presented a prototype of such a framework for solving the problem of community detection in networks on two distinctively different architectures: D-Wave quantum annealer and universal quantum computer. We suggest extending this approach for solving other types of problems in science.

The constant change of hardware and overall immaturity of the existing technology leads to many risks in QC. In spite of major effort, it has not been experimentally demonstrated yet an ability to achieve speedups over state-of-the-art classical supercomputers and there are valid concerns about scalability of existing implementations~\cite{brugger2018quantum}. Advances in material design and engineering will allow the community to overcome those hurdles. We expect QC to eventually become a part of the HPC ecosystem with an initial role as an accelerator providing a new layer of parallelism. Our approach will provide for co-design exploration towards the best QC accelerator choice for an application mix.

\clearpage
\appendices

\section*{Acknowledgment}
This research used the resources of the Argonne Leadership Computing Facility, which is a U.S. Department of Energy (DOE) Office of Science User Facility supported under Contract DE-AC02-06CH11357. We gratefully acknowledge the computing resources provided and operated by the Joint Laboratory for System Evaluation (JLSE) at Argonne National Laboratory.  The authors would also like to acknowledge the NNSA’s Advanced Simulation and Computing (ASC) program at Los Alamos National Laboratory (LANL) for use of their Ising D-Wave 2X quantum computing resource and D-Wave Systems Inc. for use of their 2000Q resource. The LANL research contribution has been funded by LANL Laboratory Directed Research and Development (LDRD).  LANL is operated by Los Alamos National Security, LLC, for the National Nuclear Security Administration of the U.S. DOE under Contract DE-AC52-06NA25396. Clemson University is acknowledged for generous allotment of compute time on Palmetto cluster.



\bibliographystyle{./IEEEtranS}
\bibliography{./refs,./qaoa}

\begin{thebibliography}{10}
\providecommand{\url}[1]{#1}
\csname url@samestyle\endcsname
\providecommand{\newblock}{\relax}
\providecommand{\bibinfo}[2]{#2}
\providecommand{\BIBentrySTDinterwordspacing}{\spaceskip=0pt\relax}
\providecommand{\BIBentryALTinterwordstretchfactor}{4}
\providecommand{\BIBentryALTinterwordspacing}{\spaceskip=\fontdimen2\font plus
\BIBentryALTinterwordstretchfactor\fontdimen3\font minus
  \fontdimen4\font\relax}
\providecommand{\BIBforeignlanguage}[2]{{%
\expandafter\ifx\csname l@#1\endcsname\relax
\typeout{** WARNING: IEEEtranS.bst: No hyphenation pattern has been}%
\typeout{** loaded for the language `#1'. Using the pattern for}%
\typeout{** the default language instead.}%
\else
\language=\csname l@#1\endcsname
\fi
#2}}
\providecommand{\BIBdecl}{\relax}
\BIBdecl

\bibitem{ambainis2018quantum}
A.~Ambainis, K.~Balodis \emph{et~al.}, ``Quantum speedups for exponential-time
  dynamic programming algorithms,'' \emph{arXiv preprint arXiv:1807.05209},
  2018.

\bibitem{amin2004}
M.~H.~S. Amin and A.~Y. Smirnov, ``Quasiparticle decoherence in {D-Wave}
  superconducting qubits,'' \emph{Phys. Rev. Lett.}, vol.~92, p. 017001, 2004.

\bibitem{bardella2016hierarchical}
G.~Bardella, A.~Bifone \emph{et~al.}, ``Hierarchical organization of functional
  connectivity in the mouse brain: a complex network approach,''
  \emph{Scientific reports}, vol.~6, p. 32060, 2016.

\bibitem{brandes2006maximizing}
U.~Brandes, D.~Delling \emph{et~al.}, ``Maximizing modularity is hard,''
  \emph{arXiv preprint physics/0608255}, 2006.

\bibitem{brugger2018quantum}
J.~Brugger, C.~Seidel \emph{et~al.}, ``Quantum annealing with disorder,''
  \emph{arXiv preprint arXiv:1808.06817}, 2018.

\bibitem{dwave2018}
\BIBentryALTinterwordspacing
{D-Wave Systems Inc.}, ``Introduction to the {D-Wave} quantum hardware,'' 2018.
  [Online]. Available:
  \url{www.dwavesys.com/tutorials/background-reading-series/introduction-d-wave-quantum-hardware}
\BIBentrySTDinterwordspacing

\bibitem{dunjko2018computational}
V.~Dunjko, Y.~Ge, and J.~I. Cirac, ``Computational speedups using small quantum
  devices,'' \emph{arXiv preprint arXiv:1807.08970}, 2018.

\bibitem{farhi2014quantum}
E.~Farhi, J.~Goldstone, and S.~Gutmann, ``A quantum approximate optimization
  algorithm,'' \emph{arXiv preprint arXiv:1411.4028}, 2014.

\bibitem{farhi2016quantum}
E.~Farhi and A.~W. Harrow, ``Quantum supremacy through the quantum approximate
  optimization algorithm,'' \emph{arXiv preprint arXiv:1602.07674}, 2016.

\bibitem{hagberg2008}
A.~A. Hagberg, D.~A. Schult, and P.~J. Swart, ``Exploring network structure,
  dynamics, and function using networkx,'' in \emph{Proceedings of the 7th
  Python in Science Conference (SciPy 2008)}, G.~Varoquaux, T.~Vaught, and
  J.~Millman, Eds., Pasadena, CA USA, 2008, pp. 11--15.

\bibitem{hager2018multilevel}
W.~W. Hager, J.~T. Hungerford, and I.~Safro, ``A multilevel bilinear
  programming algorithm for the vertex separator problem,'' \emph{Computational
  Optimization and Applications}, vol.~69, no.~1, pp. 189--223, 2018.

\bibitem{harris2018}
R.~Harris, Y.~Sato \emph{et~al.}, ``Phase transitions in a programmable quantum
  spin glass simulator,'' \emph{Science}, vol. 361, pp. 162--165, 2017.

\bibitem{king2018observation}
A.~D. King, J.~Carrasquilla \emph{et~al.}, ``Observation of topological
  phenomena in a programmable lattice of 1,800 qubits,'' \emph{arXiv preprint
  arXiv:1803.02047}, 2018.

\bibitem{kunegis2013konect}
J.~Kunegis, ``Konect: the koblenz network collection,'' in \emph{Proceedings of
  the 22nd International Conference on World Wide Web}.\hskip 1em plus 0.5em
  minus 0.4em\relax ACM, 2013, pp. 1343--1350.

\bibitem{leyffer2013fast}
S.~Leyffer and I.~Safro, ``Fast response to infection spread and cyber attacks
  on large-scale networks,'' \emph{Journal of Complex Networks}, vol.~1, no.~2,
  pp. 183--199, 2013.

\bibitem{mcclean2016theory}
J.~R. McClean, J.~Romero \emph{et~al.}, ``The theory of variational hybrid
  quantum-classical algorithms,'' \emph{New Journal of Physics}, vol.~18,
  no.~2, p. 023023, 2016.

\bibitem{mcgeoch2014adiabatic}
C.~C. McGeoch, ``Adiabatic quantum computation and quantum annealing: Theory
  and practice,'' \emph{Synthesis Lectures on Quantum Computing}, vol.~5,
  no.~2, pp. 1--93, 2014.

\bibitem{neukart2017}
F.~Neukart, G.~Compostella \emph{et~al.}, ``Traffic flow optimization using a
  quantum annealer,'' \emph{Frontiers in ICT}, vol.~4, pp. 1--6, 2017.

\bibitem{2006PNAS..103.8577N}
M.~E.~J. {Newman}, ``{From the Cover: Modularity and community structure in
  networks},'' \emph{Proceedings of the National Academy of Science}, vol. 103,
  pp. 8577--8582, Jun. 2006.

\bibitem{nicolini2017community}
C.~Nicolini, C.~Bordier, and A.~Bifone, ``Community detection in weighted brain
  connectivity networks beyond the resolution limit,'' \emph{Neuroimage}, vol.
  146, pp. 28--39, 2017.

\bibitem{nielsen2002quantum}
M.~A. Nielsen and I.~Chuang, ``Quantum computation and quantum information,''
  2002.

\bibitem{omalley2017}
D.~O'Malley, V.~V. Vesselinov \emph{et~al.}, ``Nonnegative/binary matrix
  factorization with a {D-Wave} quantum annealer,'' \emph{arXiv preprint
  arXiv:1704.01605}, 2017.

\bibitem{optimization2014inc}
G.~Optimization, ``Inc.,“gurobi optimizer reference manual,” 2015,''
  \emph{URL: http://www. gurobi. com}, 2014.

\bibitem{otterbach2017unsupervised}
J.~Otterbach, R.~Manenti \emph{et~al.}, ``Unsupervised machine learning on a
  hybrid quantum computer,'' \emph{arXiv preprint arXiv:1712.05771}, 2017.

\bibitem{palla2005uncovering}
G.~Palla, I.~Der{\'e}nyi \emph{et~al.}, ``Uncovering the overlapping community
  structure of complex networks in nature and society,'' \emph{Nature}, vol.
  435, no. 7043, p. 814, 2005.

\bibitem{preskill2018quantum}
J.~Preskill, ``Quantum computing in the nisq era and beyond,'' \emph{arXiv
  preprint arXiv:1801.00862}, 2018.

\bibitem{romero2018strategies}
J.~Romero, R.~Babbush \emph{et~al.}, ``Strategies for quantum computing
  molecular energies using the unitary coupled cluster ansatz,'' \emph{Quantum
  Science and Technology}, 2018.

\bibitem{rotta2011multilevel}
R.~Rotta and A.~Noack, ``Multilevel local search algorithms for modularity
  clustering,'' \emph{Journal of Experimental Algorithmics (JEA)}, vol.~16, pp.
  2--3, 2011.

\bibitem{smelyanskiy2016qhipster}
M.~Smelyanskiy, N.~P. Sawaya, and A.~Aspuru-Guzik, ``qhipster: the quantum high
  performance software testing environment,'' \emph{arXiv preprint
  arXiv:1601.07195}, 2016.

\bibitem{su2010glay}
G.~Su, A.~Kuchinsky \emph{et~al.}, ``Glay: community structure analysis of
  biological networks,'' \emph{Bioinformatics}, vol.~26, no.~24, pp.
  3135--3137, 2010.

\bibitem{ushijima2017graph}
H.~Ushijima-Mwesigwa, C.~F. Negre, and S.~M. Mniszewski, ``Graph partitioning
  using quantum annealing on the d-wave system,'' in \emph{Proceedings of the
  Second International Workshop on Post Moores Era Supercomputing}.\hskip 1em
  plus 0.5em minus 0.4em\relax ACM, 2017, pp. 22--29.

\bibitem{yang2017optimizing}
Z.-C. Yang, A.~Rahmani \emph{et~al.}, ``Optimizing variational quantum
  algorithms using pontryagin's minimum principle,'' \emph{Physical Review X},
  vol.~7, no.~2, p. 021027, 2017.

\end{thebibliography}

\end{document}